\input harvmac
\input epsf
\newcount\figno
\figno=0
\def\fig#1#2#3{
\par\begingroup\parindent=0pt\leftskip=1cm\rightskip=1cm\parindent=0pt
\baselineskip=11pt
\global\advance\figno by 1
\midinsert
\epsfxsize=#3
\centerline{\epsfbox{#2}}
\vskip 12pt
{\bf Fig.\ \the\figno: } #1\par
\endinsert\endgroup\par
}
\def\figlabel#1{\xdef#1{\the\figno}}
\def\encadremath#1{\vbox{\hrule\hbox{\vrule\kern8pt\vbox{\kern8pt
\hbox{$\displaystyle #1$}\kern8pt}
\kern8pt\vrule}\hrule}}

\lref\rDMJR{``{\it Collective String Field Theory of Matrix Models in the BMN
Limit}", R. de Mello Koch, A. Jevicki and J. P. Rodrigues, hep-th/0209155.}

\lref\rBMN{``{\it Strings in Flat Space and pp Waves from N=4 Super
Yang-Mills}",
D. Berenstein, J. Maldacena and H. Nastase, {\it JHEP} {\bf 0204}:013, (2002),
hep-th/0202021.}

\lref\rM{``{\it The Large $N$ Limit of Superconformal Field Theories and
Supergravity}",J.M. Maldacena, Adv. Theor. Math. Phys. {\bf 2} 231, (1998), 
hep-th/9711200.}

\lref\rGKP{``{\it Gauge Theory Correlators from Noncritical String Theory}",
S.S. Gubser, I.R. Klebanov and A.M. Polyakov, Phys. Lett. {\bf B428} 105
(1998).}

\lref\rW{``{\it Anti-de Sitter Space and Holography,}" E. Witten, Adv. Theor.
Math. Phys.{\bf 2} 253 (1998), hep-th/9802150.}

\lref\rY{``{\it What is Holography in the plane-wave limit of AdS/CFT
Correspondence}", T. Yoneya,
hep-th/0304183.}

\lref\rSFT{See for example
``{\it Explicit formulas for Neumann coefficients in the plane wave geometry,}"
Y.H. He, J. H. Schwarz, M. Spradlin and A. Volovich, Phys. Rev. {\bf D67}:086005,
2003, hep-th/0211198\semi
{\it ``An Alternative Formulation of Light Cone String Field Theory on the Plane 
wave,}"  A.Pankiewicz, hep-th/0304232.}

\lref\rFM{``{\it pp Wave String Interactions from Perturbative Yang-Mills 
Theory}," N. R. 
Constable, D Freedman, M. Headrick, S. Minwalla, L. Motl, A. Postnikov,  W. 
Skiba {\it JHEP}
{\bf 0207}:017, (2002), hep-th/0205089\semi
{\it ``Three Point Functions in N=4 Yang-Mills Theory and pp Waves,}"
Chong-Sun Chu, V. V. Khoze and G. Travaglini, JHEP {\bf 0206:011} (2002) 
hep-th/0206005.} 

\lref\rSbook{``{\it Quantum Theory of Many Variable Systems and Fields},"
B. Sakita, World Sci. Lect. Notes Phys.1:1-217 (1985).}

\lref\rBKPS{{\it ``BMN Correlators and Operator Mixing in N=4 SYM Theory,}" N. Beisert,
C. Kristjansen, J. Plefka, G.W. Semenoff and M. Staudacher, Nucl.Phys. {\bf B650} 125-161 
(2003), hep-th/0208178.} 

\lref\rCFHM{``{\it Operator Mixing and the BMN Correspondence}," N. Constable, 
D.Z. Freedman, M. Headrick and S.Minwalla, hep-th/0209002.}

\lref\rPSVVV{{\it ``Tracing the String: BMN Correspondence at Finite $J^2/N$,}" J. Pearson, 
M. Spradlin, D. Vaman, H. Verlinde and A. Volovich, 
hep-th/0210102.} 

\lref\rGMP{{\it ``SYM Description of SFT Hamiltonian in a pp wave Background,}" J. Gomis, 
S. Moriyama and Jong-won Park, hep-th/0210153\semi
{\it ``SYM Description of pp Wave String Interactions: Singlet Sector
and Arbitrary Impurities,}" J. Gomis, S. Moriyama and Jong-won Park, 
hep-th/0301250.}

\lref\rGMP{{\it ``SYM Description of pp Wave String Interactions: Singlet Sector
and Arbitrary Impurities,}" J. Gomis, S. Moriyama and Jong-won Park, 
hep-th/0301250.}


\Title{ \vbox {\baselineskip 12pt\hbox{}
\hbox{}  \hbox{May 2003}}}
{\vbox {\centerline{ Derivation of String Field Theory from}
\centerline{the Large $N$ BMN Limit}
}}

\smallskip
\centerline{Robert de Mello Koch$^\dagger$, Aristomenis Donos$^*$, Antal
Jevicki$^*$ and Jo\~ao P. Rodrigues$^\dagger$}
\smallskip
\centerline{\it Department of Physics and Center for Theoretical
Physics$^\dagger$,}
\centerline{\it University of the Witwatersrand,}
\centerline{\it Wits, 2050, South Africa}
\centerline{\tt robert,joao@neo.phys.wits.ac.za}
\smallskip
\centerline{\it Department of Physics$^*$,}
\centerline{\it Brown University,}
\centerline{\it Providence, RI 02912, USA}
\centerline{\tt donos,antal@het.brown.edu}
\bigskip

{\vskip 20pt}

\noindent
We continue the development of a systematic procedure for deriving closed string
pp wave string field theory from the large $N$ Berenstein-Maldacena-Nastase
limit.
In the present paper the effects of the Yang-Mills interaction are considered in
detail for general BMN states. The SFT interaction with the appropriate
operator insertion at the interaction point is demonstrated.


\Date{}

\def\MAKEdalamSIGN#1#2{%
\setbox0=\hbox{$\mathsurround 0pt #1{#2}$}
\dimen0=\ht0 \advance\dimen0 -0.8pt
\hbox{\vrule\vbox to\ht0{\hrule width\dimen0 \vfil\hrule}\vrule}}


\newsec{Introduction}
The Berenstein-Maldacena-Nastase correspondence outlines a precise
relationship between large N  ${\cal N}=4$ super Yang-Mills
theory and closed string theory in the ppwave background\rBMN. This limit
simplifies and extends the AdS/CFT correspondence\rM,\rGKP,\rW\ and is a subject of
detailed studies. Of particular interest is the derivation of pp wave string
field theory from the Yang-Mills/matrix theory.

In the present paper we continue the work begun in \rDMJR\ (here after referred
to as (I)) on developing a direct and systematic approach for mapping large $N$
gauge theory to closed string theory. We have applied in I the methods
of collective field theory and adapted them to
the large $N$ BMN limit. The approach was described in the context of a quantum
mechanical model.
The essential feature of the collective field approach is the
mechanism of joining and splitting (of loops) represented in an effective
hamiltonian. Consequently the simplest example
of obtaining string type interactions was seen to appear already from a free
matrix theory.
It was shown in  particular  that the SUGRA type amplitudes
are correctly reproduced through the above mechanism.
Likewise the 3-string overlap vertex was seen to naturally arise.
In this approach the effect
of Yang-Mills interactions is expected to renormalize the zeroth order
expressions and we discuss  this in the present work.

The matrix model language was used essentially for notational
simplicity. The matrix model contains the effects,
which in our framework, are identical in the
full gauge theory. It may also be that there is an effective matrix model
description
of the full pp theory and  our result have shed light on this possibility.
(For a most recent discussion of holography in the pp limit see\rY.)

Our plan is as follows:
In Sect.2 we summarize the basics of the approach reviewing and extending
the methods of I. In Sect.3 we give a simple example of string type
amplitudes and exhibit the
manner in which Yang-Mills type interactions modify the basic couplings.
We make use of a coherent
state picture and describe a map between this and the physical picture of I.
In Sect.4 we then take these interactions into account
concentrating on general string-type states.
In the continuum BMN limit we
derive the emerging 3-string interaction with the appropriate operator
prefactor. We reproduce the spectrum of BMN loops in an appendix.

\newsec{Collective field theory}

In this section, we give a general overview of collective field theory
as applied to the large N limit of Berenstein, Maldacena and Nastase. This
section defines the
formalism and gives a short  summary of the results reached in I. The collective
method in general provides a systematic formalism for describing the dynamics of
observable, physical degrees of
freedom in a theory with large $N$ symmetry.
In gauge or matrix theory the physical observables are given by
loops or traces of matrix products (words).
The method then provides a direct change of variables
to the invariant observables.
The resulting effective or collective Hamiltonian
describes the full dynamics of these invariants.
The essential two terms that define the
effective hamiltonian are the interaction terms containing joining and splitting
of the loops. In ${\cal N}$=4 SUSY Yang-Mills gauge theory we follow the degrees
of
freedom consisting of the Higgs fields:
 $\phi_1 \phi_2 \cdot\cdot\cdot\,\phi_5, \phi_6$. In the proposal of
Berenstein, Maldacena and Nastase two of the Higgs matrices
$(\phi_5,\phi_6 )\,\,$ are chosen to play a special
role , they are selected to define the light cone coordinates
$Z = \phi_5 + i\phi_6 $.
The above matrix variables define a sector of the full theory that we are
interested in.
In addition we consider these as functions of time only, so that the dynamics
reduces to matrix quantum mechanics. This gives a minimal set of fields that are
capable of capturing the essential features of  the BMN correspondence.

We therefore concentrate on a complex multi-matrix system with a Hamiltonian
that is invariant under

$$ Z_i \to U^{\dagger} Z_i U \qquad
\bar{Z}_i \to U^{\dagger} \bar{Z}_i U .$$

\noindent
The basic equal time variables are given by single trace operators

\eqn\Word{\Tr \left(...
\prod_{i=1}^M Z_i^{n_i} {\bar{Z}_i}^{\bar{n}_i}
\prod_{j=1}^M Z_j^{m_j} {\bar{Z}_j}^{\bar{m}_j}
...\right).}

\noindent
In the large $N$ limit, one then has a change of variables from the original
Yang-Mills fields to the invariant loop variables denoted collectively by
$\phi_{C}$,
where ${C}$ stands for a loop or word index. This index also includes complex
conjugate loops $\bar{\phi}_{C}$. One then considers a change of variables to
this new set in the canonical operator formalism. Concentrating on the kinetic
term, there follows

\eqn\Kinetic{
T = - \Tr\left(\sum_{i=1}^M {\partial\over\partial \bar{Z}_i}
{\partial\over\partial {Z}_i}\right) =
-\sum_{C,C'} \Omega (C,C') {\partial\over\partial \bar{\phi}_C}
{\partial\over\partial {\phi}_{C'}} +
\sum_{C}  \omega (C) {\partial\over\partial {\phi}_{C}}
}

\noindent
with

$$  \Omega (C,C') = \Tr\left(\sum_{i=1}^M
{\partial\bar{\phi}_C\over\partial\bar{Z}_i}{\partial{\phi}_{C'}\over\partial
{Z}_i}\right)
= \bar{\Omega}(C',C)
$$

\noindent
and

$$
\omega (C) = - \Tr\left(\sum_{i=1}^M {\partial^2\phi_{C}\over\partial
\bar{Z}_i\partial {Z}_i}\right).
$$

\noindent
These two operations define the processes of joining and splitting. In
particular
$\Omega (C,C')$ ``joins" loops, or words. As an example, if
$\phi_{C} = \Tr (Z_1^J)$ and $\phi_{C'} = \Tr (Z_1^{J'})$ then
$\Omega = J {J'} \Tr (Z_1^{J-1} \bar Z_1^{J'-1})$. In general, one has
schematically

$$   \Omega (C,C') =  \sum \phi_{C+C'}   $$

\noindent
where ${C+C'}$ is obtained by adding the two words $C$ and $C'$.
Similarly, $\omega$ ``splits" loops. Again,

\eqn\littleo{\omega (C) = \sum  \phi_{C'}\phi_{C''} }

\noindent
represents the processes of splitting the word $C$ into
$C'$ and $C''$. The hermiticity of the new collective
description is assured through a field transformation
and this completes the hamiltonian\rSbook.

The non-triviality in applying collective field theory to multimatrix models
comes from the enormous set of loops (words) that can be generated. For
the present problem Berenstein,Maldacena and Nastase have identified a set
of observables (traces)
which have a mapping into the pp wave string. If
we denote the BMN
set of loops by $C$ and their conjugates by $C'$, the gauge theory process of
``joining"
(contained in $\Omega (C,C')$) generates
new loops not in the original set. It the question of extra
degrees of freedom that was first addressed and clarified in I.
Consider the collective variables that have a direct relation with (lattice)
string fields

\eqn\LattStr
{\Phi_J(\{ l\})=\Tr\Big( T_l\prod_{i=1}^n\phi (l_i) Z^J\Big),\qquad
\phi (l_i )=Z^{l_i}\phi Z^{-l_i},}

\noindent
with $T_l$ the $l$ ordering operator - it orders the $\phi (l)$ factors so that
$l_i$ increases from left to right. We also have

\eqn\ConjFld
{\bar{\Phi}_J (\{ l\})=\Tr\Big( \bar{Z}^J\tilde{T}_l\prod_{i=1}^n \bar{\phi}
(l_i) \Big),\qquad
\bar{\phi} (l_i )=\bar{Z}^{-l_i}\bar{\phi}\bar{Z}^{l_i}.}

\noindent
$\tilde{T}_l$ is a second $l$ ordering operator - it orders the $\bar{\phi}(l)$
factors so that $l_i$ {\it decreases} from left to right.

The joining and splitting processes applied to these lattice string fields
proceed as follows. One first has the open loop defined by

$$ P_{J}(\{ l\}))_{ij}=\sum_{a=1}^J T_l\Big(\Big[\prod_{i=1}^n \phi
(l_i-a)Z^{J-1}\Big]_{ij}\Big)
= {\partial \Phi_J(\{ l\})\over\partial Z_{ji}}$$

\noindent
where we have $l_i-a$ mod $J$ so that $0\le l_i-a\le J-1$. Similarly

$$ Q_{J}(\{ l\}))_{ij}=\sum_{j=1}^n T_l\Big(\prod_{i=1,i\ne j}^n
\Big[\phi (l_i-l_j)Z^{J}\Big]_{ij}\Big)=
{\partial \Phi_J(\{ l\})\over\partial \phi_{ji}}$$

\noindent
In terms of these split (open) loops we generate through the joining
operation the composite, joined loop.
In particular these are contained in the interaction of the collective
hamiltonian through the composite trace $\Omega$

$$\eqalign{ {\partial\Phi_{J_1}(\{ l\})\over\partial Z_{ij}}
&{\partial\bar{\Phi}_{J_2}(\{ \bar{l}\})\over\partial\bar{Z}_{ji}}
+ {\partial\Phi_{J_1}(\{ l\})\over\partial \phi_{ij}}
{\partial\bar{\Phi}_{J_2}(\{ \bar{l}\})\over\partial\bar{\phi}_{ji}}\cr
&=\Tr (P_{J_1}(\{ l\})\bar{P}_{J_2}(\{\bar{l}\})) +\Tr (Q_{J_1}(\{
l\})\bar{Q}_{J_2}(\{\bar{l}\}))\cr
&=\sum_{a=1}^{J_1}\sum_{b=1}^{J_2}\Tr\Big[ T_l\Big(\prod_{i=1}^{n_1}\phi
(l_i-a)\Big)Z^{J_1-1}
\bar{Z}^{J_2-1}\tilde{T}_l\Big(\prod_{j=1}^{n_2}\bar{\phi}(\bar{l}_j-b)\Big)\Big
]\cr
&+\sum_{k=1}^{n_1}\sum_{m=1}^{n_2}\Tr\Big[ T_l\Big(\prod_{i=1,i\ne k}^{n_1}\phi
(l_i-l_k)\Big)Z^{J_1}
\bar{Z}^{J_2}\tilde{T}_l\Big(\prod_{j=1,j\ne
m}^{n_2}\bar{\phi}(\bar{l}_j-\bar{l}_m)\Big)\Big]}$$

\noindent
This composite trace involves mixed combinations of Z and $\bar{Z}$ which do
not survive as excitations in the BMN correspondence. In I we have introduced
a mechanism of factorization based on which we replace such new, unwanted loops
by a sum of physical, BMN approved loops. Concretely,

$$
Tr \left( P_{J_{1}} ( \{ l\} ) \, \bar{P}_{J_{2}}
\,(\{\bar{l}\})\right) = \sum_{J,\{m\}} \, C_{J J_1
J_2}^{\{m\}\{l\}\{\bar{l}\}}\, \bar{\Phi}_J \left( \{m\} \right)
$$

\noindent
The coefficients (structure constants) in this relation can be generated by
Schwinger-Dyson
equations (as we have discussed in (I)). A more complete procedure for finding
these coefficients is
through consistency conditions obtained by taking expectation values of both
sides of the equation. In
this way they can be read of from the correlator

$$
\langle \Phi_J \ (\{m\}) \, Tr \left( \, P_{J_{1}} (\{ l \} ) \,
P_{J_{2}} ( \{ \bar{l} \} ) \right) \rangle
$$
to equal

$$\eqalign{&C_{J J_{1} J_{2} }^{\{m\}\{l\} \{\bar{l}\} }=
J_{1}  \sum_{\kappa=1}^{n} \, \sum_{l = 1}^{n_{1}} \, \sum_{q=1}^{n_2} \prod_{i=1}^{n-1} \prod_{j=1}^{n_{1}-1} \delta (m_{\kappa + i\, mod\, n}
- m_{\kappa}\, mod\, J , \bar{l}_{q+i\, mod\, n_{2}} - \bar{l}_{q}\, mod\, J_{2})\cr &\quad \times \delta ( l_{l+jmodn_{1}} - l_{l} mod\,
J_{1}+\bar{l}_{q+n\, mod\, n_{2}} - \bar{l}_{q}\, mod\, J_{2}, \bar{l}_{q+n+j\, mod\, n_{2}} - \bar{l}_q\, mod\, J_{2}) \cr & \quad \times
min\{m_{\kappa}-m_{\kappa-1modn},l_{l}-l_{l-1modn_1}\},}
$$
\noindent Similarly

$$
Tr \left( Q_{J_{1}} ( \{ l \} ) \bar{Q}_{J_{2}} ( \{\bar{l}\} )
\right) = \sum_{J, \{m\}} \, G_{J J_{1} J_{2}}^{\{
m\}\{l\}\{\bar{l}\}} \, \bar{\Phi}_J \left( \{m\}\right)
$$

\noindent
with

$$\eqalign{
&G_{J J_{1} J_{2}}^{\{ m\} \{ l\}\{ \bar{l}\}}= n_{1} \sum_{\kappa=1}^{n} \, \sum_{l = 1}^{n_{1}} \, \sum_{q=1}^{n_2} \prod_{i=1}^{n-1}
\prod_{j=1}^{n_{1}-1} \delta (m_{\kappa + i\, mod\, n} - m_{\kappa}\, mod\, J , \bar{l}_{q+i\, mod\, n_{2}} - \bar{l}_{q}\, mod\, J_{2})\cr
&\quad \times \delta ( l_{l+jmodn_{1}} - l_{l} mod\, J_{1}+\bar{l}_{q+n\, mod\, n_{2}} - \bar{l}_{q}\, mod\, J_{2}, \bar{l}_{q+n+j\, mod\,
n_{2}} - \bar{l}_q\, mod\, J_{2}) \cr & \quad \times min\{m_{\kappa}-m_{\kappa-1modn},l_{l}-l_{l-1modn_1}\},}
$$

\noindent
is again most simply deduced from the correlator

$$
\langle \Phi_J \ (\{m\}) \, Tr \left( \, Q_{J_{1}} (\{ l \} ) \,
Q_{J_{2}} ( \{ \bar{l} \} ) \right) \rangle
$$

\noindent
The above  sums appearing in $C$ and $G$ are equal and represent a form of
a (lattice) three-string
vertex  $|V_{3}^{0}\rangle$. In order to demonstrate that, let us consider the
string states

$$
|\psi_{1}\rangle= \sum_{p=0}^{J_{1}-1} \prod_{i=1}^{n_{1}} b_{p+ l_{i}\, mod\,
J_{1}}^{\left(1\right) \, \dagger} |0\rangle_{1}
$$
$$
|\psi_{2}\rangle= \sum_{q=0}^{J-1} \prod_{j=1}^{n} b_{q+ m_{j}\, mod\,
J}^{\left(2\right) \, \dagger} |0\rangle_{2}
$$
$$
|\psi_{3}\rangle= \sum_{r=0}^{J_{2}-1} \prod_{k=1}^{n_{1}} b_{r+ \bar{l}_k\,
mod\, J_{1}}^{\left(3\right) \, \dagger} |0\rangle_{3}
$$

\noindent
these are in direct correspondence with our matrix theory fields (states)
$\bar{\Phi}_J \left( \{m\}\right)$, $\Phi_{J_1} \left( \{l\}\right)$ and
$\bar{\Phi}_{J_2} \left(\{\bar{l}\}\right)$ .

The sums appearing in the matrix theory result correspond to reparametrizations that occur in the calculation of $\langle\psi_{1}| \,
\langle\psi_{2}| \, \langle\psi_{3}| \, |V_{3}^{0}\rangle$. The sum in $C$ would appear in the above calculation if we fix one of the delta
functions between string (2) and (3) and one of the delta functions between (1) and (3). This allowed us to write the part of our collective
Hamiltonian coming from $H_0$ as
$$
\eqalign{ & H_0^{col}= \sum_{J,\{l\}} (J+n) \Phi_J(\{l\}) {\partial\over
\partial \Phi_J(\{l\})}+ \cr & \sum_{J_1,J_2,J_3}
\sum_{\{l^{(1)}\},\{l^{(2)}\},\{l^{(3)}\}} \left(\Delta J+\Delta n \right)
\langle\psi_{1}| \, \langle\psi_{2}| \, \langle\psi_{3}|
\,
|V_{3}^{0}\rangle \Phi_{J_3}(\{l^{(3)}\}) {\partial\over \partial
\Phi_{J_1}(\{l^{(1)}\})} {\partial\over \partial
\Phi_{J_2}(\{l^{(2)}\})} \cr
& +\sum_{J_1,J_2,J_3} \sum_{\{l^{(1)}\},\{l^{(2)}\},\{l^{(3)}\}} \left(\Delta
J+\Delta n \right) \langle\psi_{1}| \,
\langle\psi_{2}| \,
\langle\psi_{3}| \, |V_{3}^{0}\rangle \Phi_{J_1}(\{l^{(1)}\})
\Phi_{J_2}(\{l^{(2)}\}) {\partial\over \partial
\Phi_{J_3}(\{l^{(3)}\})}}
$$
\noindent where
$$
\Delta J+\Delta n=J_1+n_1+J_2+n_2-J_3-n_3
$$
\noindent
and $|\psi_{i} \rangle$ are the lattice string states associated with
$\Phi_{J_i}(\{l^{i}\})$.
The characteristic feature of this interaction is that it appears proportional
to $1/N$ and
exhibits the prefactor
$(E_3^0-E_1^0-E_2^0)$. The occurrence of the energy prefactor was an early
conjecture of\rFM. It was explained
in I in a specific example as resulting from  a
projection to light cone fields.
It is expected that this form will be
modified once Yang-Mills interactions are taken into account.

We note that this interaction can be transformed away through a (nonlinear)
field redefinition.
The Hamiltonian $H_2 + H_3$ can be reduced (in leading order of $1/N$) to $H_2$.
Another way to understand this fact is to realize that in the
creation-annihilation (coherent state) basis
the free oscillator hamiltonian is first order in the derivatives and its
collective representation in this basis
is still a quadratic Hamiltonian
$$
H_2 = \sum \, E_i A^{\dagger} (\{ l_i\} ) \, A ( \{l_i \} ).
$$

\noindent
The nonlinear transformation

$$
\eqalign{A^{\dagger} ( \{ l_i \} ) &= \Phi^{\dagger} (\{ l_i \} ) + {1\over 4N}
\bar{C}
(\{ l_i \} , \{ l_j\}\{ l_k\} )
\Phi^{\dagger} (\{ l_j \} ) \, \Phi^{\dagger} (\{ l_k \} )\cr
&+ {1\over 2N} \, C (\{ l_j \} , \{ l_i \} \{ l_k \} ) \,
\Phi (\{ l_k\} ) \Phi^{\dagger} (\{ l_j\} )}
$$

\noindent
from coherent state fields $\, A (\{ l\} )\,$ to the physical collective fields
relates the
two representations. The coherent state
picture will in the next section provide the simplest framework for
incorporating the
effect of Yang-Mills
interactions.

\newsec{Examples}

We will start by considering the $g_{YM}^2$ effects of the dimensionally reduced
Yang-Mills system and discuss simple, illustrative examples of corrections that
this term gives. Consider the Hamiltonian

$$H = \sum_{i=1}^6 \Tr \left( - {\partial^2\over \partial\phi_i^2} + \phi_i^2
\right) \,-g_{YM}^2\sum_{i<j}\Tr\left(\big[\phi_i,\phi_j\big]^2\right),\qquad
i,j=1,...,6 .$$

\noindent
With

$$\phi=\phi_1+i\phi_2,\qquad \psi=\phi_3+i\phi_4,\quad Z=\phi_5+i\phi_6,$$

\noindent
the interaction term in the above Hamiltonian is equivalently written as

\eqn\Interaction {\eqalign{H_1=g_{YM}^2\Big(&{1\over
4}\big[\phi,\bar{\phi}\big]\big[\phi,\bar{\phi}\big] +{1\over
4}\big[\psi,\bar{\psi}\big]\big[\psi,\bar{\psi}\big] +{1\over
4}\big[Z,\bar{Z}\big]\big[Z,\bar{Z}\big]\cr
&+{1\over
2}\big[\phi,\bar{\phi}\big]\big[\psi,\bar{\psi}\big] +{1\over
2}\big[\phi,\bar{\phi}\big]\big[Z,\bar{Z}\big] +{1\over
2}\big[\psi,\bar{\psi}\big]\big[Z,\bar{Z}\big]\cr
&-\big[\bar{\phi},\bar{Z}\big]\big[\phi,Z\big]
-\big[\bar{\psi},\bar{Z}\big]\big[\psi,Z\big]
-\big[\bar{\phi},\bar{\psi}\big]\big[\phi,\psi\big]\Big),}}

\noindent
showing the usual split into $D$ and $F$ terms respectively. Having in mind the
passage to the infinite momentum frame, we work in a coherent state basis and
project

\eqn\Project
{\bar{Z}\to A^\dagger +B\to A^\dagger,\qquad
Z=A+B^\dagger\to {\partial\over\partial A^\dagger}.}

\noindent
We don't consider $B^\dagger$ quanta. These quanta correspond, in the pp-wave
string field theory, to modes with $p^+<0$. In the infinite momentum frame
these modes decouple.

For the complex impurity fields $\psi$ and $\phi$ we project

\eqn\ProjectImp
{\eqalign{\bar{\phi}&\to b^\dagger +d\to b^\dagger,\qquad
\phi \to b +d^\dagger\to {\partial\over\partial b^\dagger},\cr
\bar{\psi}&\to c^\dagger +e\to c^\dagger,\qquad
\psi \to c +e^\dagger\to {\partial\over\partial c^\dagger}.}}

\noindent
This last projection ensures that our hamiltonian keeps us within the subspace
of
loops that are near to chiral primary operators. This truncation could be
justified
by appealing to supersymmetry. Here we follow a more pedestrian approach and
justify
the truncation by showing that the resulting hamiltonian reproduces the impurity
number conserving amplitudes of the light cone string field theory vertex.
We have also checked that the correct
string mass spectrum (i.e. anomalous dimensions) is obtained to order $1/N^2$.

An immediate consequence of the above reduction of degrees of freedom is that
the $D$ terms in \Interaction\ are trivial as a result of the commutator

\eqn\DTriv
{\left[\left({\partial\over\partial
A^\dagger}\right)_{ij},A^\dagger_{kl}\right]=
\left[ {\partial\over\partial A^\dagger_{ji}},A^\dagger_{kl}\right]=
\delta_{jk}\delta_{il}.}

\noindent
We therefore arrive at the Hamiltonian

\eqn\HamFnl
{\hat{H}=-g_{YM}^2\Big(
\big[b^\dagger ,A^\dagger\big]\big[{\partial\over\partial b^\dagger}
,{\partial\over\partial A^\dagger}\big]
+\big[c^\dagger ,A^\dagger\big]\big[{\partial\over\partial c^\dagger}
,{\partial\over\partial A^\dagger}\big]
+\big[b^\dagger ,c^\dagger\big]\big[{\partial\over\partial b^\dagger}
,{\partial\over\partial c^\dagger}\big]\Big)}

\noindent
which can be recognized as the (dimensionally reduced) operator
$\hat{\Delta}-\hat{J}$ in a coherent state basis. In obtaining this
expression, we have subtracted the free terms in the Hamiltonian, which
in the large $J$ limit simply contribute an additive term equal to $\hat{J}$.

In I, we have already demonstrated agreement for a class of sugra states given
by the loop variables

\eqn\TwoImp
{O_{n,m}^J=\sum\Tr\big(\phi^n\psi^m Z^J\big) .}

\noindent
The sum is over all possible permutations of the $\phi$ and $\psi$ fields, that
is,
the above loop is a chiral primary operator. Based on the free part of the above
hamiltonian the following interacting cubic hamiltonian was shown to arise

\eqn\TwoImp
{\eqalign{H&=2\mu\delta_{J_1,J_2+J_3}\delta_{n_1,n_2+n_3}\delta_{m_1,m_2+m_3}
{\sqrt{J_1 J_2 J_3}\over N}\sqrt{n_1!\over n_2! n_3!}
\sqrt{m_1!\over m_2! m_3!}\cr
&\times \Big({J_2\over J_1}\Big)^{n_2+m_2\over 2}
\Big({J_3\over J_1}\Big)^{n_3+m_3\over 2}
\Pi^{\prime J_1}_{n_1,m_1}
\bar{\Pi}^{\prime J_2}_{n_2,m_2}
O^{\prime J_3}_{n_3,m_3}.}}

\noindent
The sugra amplitudes are special in that they do not seem to receive corrections
from
the Yang-Mills interaction. This is clear for their energies (i.e. anomalous
dimensions) where non-renormalization theorems have been obtained. For the
3-point
couplings one can also demonstrate an absence of corrections (in the leading
order).
Here we consider the next set of operators of interest, which contain stringy
excitations and consequently do receive corrections from the Yang-Mills
interaction. Let
us concentrate on the subset of the full loop space consisting of the gauge
theory
operators

$$\eqalign{\tilde{O}^J&=\sqrt{JN^J}O^J=\Tr (A^{\dagger J}),\cr
\tilde{O}^J_n&=\sqrt{JN^J}O^J_n=\sum_{l=0}^J q^l\Tr (b^\dagger
(A^\dagger )^l c^\dagger (A^\dagger )^{J-l})=\sum_{l=0}^J q^l O_l .}$$

\noindent
where\foot{In the large $J$ limit, the difference between $J$ and $J+1$
is inconsequential. We use $J+1$ to simplify the transform between $O_l$ and
$\tilde{O}_n^J$.} $q=e^{2\pi in\over J+1}$. Begin by considering
the action of $\hat{H}$ on the two impurities state

$$\eqalign{\hat{H}\tilde{O}^J_n&=-g_{YM}^2\Big( 2N\big[
\sum_{l=1}^J q^l(O^J_{l-1}-O_l^J)+\sum_{l=0}^{J-1} q^l(O^J_{l+1}-O_l^J)\big]\cr
&+\sum_{l=2}^J
q^l\sum_{l'=1}^{l-1}\tilde{O}^{l'}(O^{J-l'}_{l-l'-1}-O_{l-l'}^{J-l'})
+\sum_{l=0}^{J-2} q^l\sum_{l'=0}^{J-l-2}\tilde{O}^{J-l-l'-1}(O^{l+l'+1}_{l+1}
-O_{l}^{l+l'+1})\cr
&+\sum_{l=2}^J
q^l\sum_{l'=0}^{l-2}\tilde{O}^{l-l'-1}(O^{J-l+l'+1}_{l'}-O^{J-l+l'+1}_{l'+1})
+\sum_{l=0}^{J-2}
q^l\sum_{l'=1}^{J-l-1}\tilde{O}^{l'}(O^{J-l'}_{l+1}-O_{l}^{J-l'})\cr
&+\sum_{l=0}^J q^l\tilde{O}^{J-l}(O^{l}_{0}-O_{l}^{l})
+\sum_{l=0}^J q^l\tilde{O}^{l}(O^{J-l}_{J-l}-O_{0}^{J-l}).}$$

\noindent
By changing the order of the double summations, expressing
$O_l=\sum_n e^{-{2\pi inl\over J+1}}\tilde{O}_n^J/(J+1)$,
performing the intermediate sums, taking into account the
normalization of the states and taking the large $J$ limit, we
obtain ($y={J_1+1\over J+1}\to {J_1\over J}$)

\eqn\HamAct
{\hat{H}|O_n^J\rangle =\lambda' 8\pi^2 n^2 |O_n^J\rangle
-g_2\lambda'\sum_{J_1+J_2=J}
\sum_{m=-{J_1\over 2}}^{J_1\over 2}{1\over\sqrt{J}}\sqrt{1-y\over y}
\Big({8m\over ny-m}\Big)\sin^2 (\pi ny)|O_m^{J_1}O^{J_2}\rangle }

\noindent
where $\lambda'$ and $g_2$ have their usual meanings

$$\lambda'={g_{YM}^2 N\over J^2},\qquad g_2={J^2\over N}.$$

\noindent
As expected, apart from the diagonal term which gives the first string tension
correction
to the anomalous dimension, $\hat{H}$ provides a splitting of  the impurity loop
into two loops.

Consider next the action of $\hat{H} $ on two loops $O_m^{J_1}O^{J_2}$. Apart
from the diagonal term and the
splitting into a 3 trace state, this will exhibit the effect of joining loops
$O_m^{J_1}$ and $O^{J_2}$ into a single BMN state $O_n^{J_1+J}$. This process is
obtained
when a derivative acts on $O_m^{J_1}$ and the other on $O^{J_2}$. Explicitly
($J=J_1+J_2$)

$$\eqalign{\hat{H}(\tilde{O}_m^{J_1}\tilde{O}^{J_2})&=-g_{YM}^2\Big[2N\Big(\sum_
{l=0}^{J_1-1}
q_m^l (O_{l+1}^{J_1}-O_l^{J_1})+\sum_{l=1}^{J_1}
q_m^l (O_{l-1}^{J_1}-O_l^{J_1})\Big)\tilde{O}^{J_2}\cr
&+2J_2\sum_{l=0}^{J_1}q_m^l\big[(O_{l+1}^{J}-O_l^{J})+
(O_{J_2+l-1}^{J}-O_{J_2+l}^{J})\big]+{\tt 3}\,\,\,{\tt trace}\,\,\,{\tt
states}\Big].}$$

\noindent
Again by re-expressing $O_l^J$ in terms of $\tilde{O}_n^J$, performing the
intermediate sums,
taking into account the normalization of the states and taking the large $J$
limit we obtain

\eqn\SecHamAct
{\eqalign{\hat{H}|O_m^{J_1}O^{J_2}\rangle
&=\lambda'\Big({8\pi^2 m^2\over y^2}\Big)|O_m^{J_1}O^{J_2}\rangle
+{\tt 3}\,\,\,{\tt trace}\,\,\,{\tt states}\cr
&-{\lambda'g_2\over\sqrt{J}}\sqrt{1-y\over y}\Big({8ny\over m-ny}\Big)
\sin^2 (\pi ny)|O_n^J\rangle .}}

There is another 2 impurity 2 trace state comprising of two loops of one
impurity each

$$ \tilde{O}^J_\phi =\sqrt{N^{J+1}}O_\phi^J=\Tr (b^\dagger A^{\dagger J}),\qquad
\tilde{O}^J_\psi =\sqrt{N^{J+1}}O_\psi^J=\Tr (c^\dagger A^{\dagger J}).$$

\noindent
Inspection of \HamAct\ shows that our Hamiltonian apparently does not split the
single 2 impurity trace into 2 single impurity traces. However, it can join the
two single impurity loops

$$\eqalign{\hat{H}(\tilde{O}^J_\phi\tilde{O}^{J_2}_\psi)&=-g_{YM}^2\Big(
\sum_{l=0}^{J_2-1}\Big[(O^J_{J-l}-O^J_{J-l-1})+(O^J_{J_2-l-1}-O^J_{J_2-l})
\Big]\cr
&+\sum_{l=0}^{J_2-1}\Big[(O^J_{l}-O^J_{l+1})+(O^J_{l+J_2+l}-O^J_{J_2+l})\Big]
\Big).}$$

\noindent
Carrying out steps similar to those leading to equations \HamAct\ and
\SecHamAct, we obtain

\eqn\ThirdHamAct
{\hat{H}|O_\phi^{J_1}O_\psi^{J_2}\rangle = -g_2\lambda'\sum_{n=-J/2}^{J/2}
{8\over\sqrt{J}}\sin^2 (\pi ny)|O_n^J\rangle +...}

\noindent
Combining the results \HamAct, \SecHamAct\ and \ThirdHamAct\ we obtain the
following Hamiltonian acting on loop space

\eqn\Summary
{\eqalign{
\hat{H} &=\lambda'(8\pi^2 n^2)O_n^J {\partial\over\partial O_n^J} \,
+\sum_{n,m,y}\lambda' g_2 D_{n,my}O_m^{J_{1}}O^{J_{2}} \, {\partial\over\partial
O_n^J}\cr
&+\sum_{n}\lambda' g_2 D_{my,n} O_n^J\, {\partial\over\partial O_m^{J_{1}}} \,
{\partial\over\partial O^{J_{2}}}\cr
&+\lambda' g_2 D_{y,n}O_n^J {\partial\over\partial O_{\phi}^{J_{1}}}
{\partial\over\partial O_{\psi}^{J_{2}}}.}}

\noindent
The coefficients can be read by inspection from equations \HamAct, \SecHamAct\
and \ThirdHamAct. In this coherent representation, we in general reach a
Hamiltonian of the form

\eqn\HCoh {H_{coh} = \sum_i \, E_i A_i^{\dagger} A_j + \sum_{ijl}D_{i,jl}\,
A_j^{\dagger} A_l^{\dagger} A_i + \sum_{ijl}F_{jl,i}
A_i^{\dagger}
A_j A_l}

\noindent
which is not hermitean. The transformation from coherent to physical fields
given in Section 2 reads

\eqn\CtoP
{A_i^{\dagger} = \psi_i^{\dagger} + {1\over 4N} \,
\bar{C}_{i,pq} \psi_p^{\dagger}
\psi_q^{\dagger} + {1\over 2N} \, C_{p,iq} \, \psi_q\psi_p^{\dagger}  .}

\noindent
It leads to an interacting Hamiltonian of the form

\eqn\HInt{\eqalign{ H &= \sum_i  E_i \psi_i^{\dagger}\psi_i + {1\over
N}\sum_{ijl} \left( (E_i - E_j - E_l ) \cdot {1\over 4}
\bar{C}_{i,jl} +
D_{i,jl} \right) \psi_j^{\dagger} \psi_l^{\dagger} \psi_i\cr &+ {1\over N}
\sum_{ijl}\left(-(E_i - E_j - E_l )\cdot {1\over 4}
C_{i,jl} +
F_{jl,i} \right) \psi_i^{\dagger} \psi_j \psi_l .}}

\noindent
For the example of three states denoted as (in what follows $y\equiv {J_1\over
J}$)

$$\eqalign{O^{J_1} & \rightarrow \psi_0^y,\cr
O_n^{J_1} & \rightarrow \psi_1^{ny},\cr
O_{\phi}^{J_1} & \rightarrow \psi_2^y,\cr
O_{\psi}^{J_1} & \rightarrow \psi_3^y,}$$

\noindent
our transformation \CtoP\ makes use of the following results

$$\eqalign{\langle O_m^{J_1}O^{J_2}|O_n^J\rangle &=g_2 C_{my,n}=g_2 C_{n,my}=
\langle O_n^J|O_{m_1}^{J_1}O^{J_2}\rangle\cr
&=g_2 {y^{3/2}\sqrt{1-y}\over\sqrt{J}\pi^2}
{\sin^2 (\pi ny)\over (m-ny)^2},}$$

$$\eqalign{\langle O_\phi^{J_1}O_\psi^{J_2}|O_n^J\rangle &=g_2 C_{y,n}=g_2
C_{n,y}=
\langle O_n^J|O_{\phi}^{J_1}O_\psi^{J_2}\rangle\cr
&=-g_2 {\sin^2 (\pi ny)\over \pi^2\sqrt{J} n^2},}$$

$$\bar{C}_{m,py}=C_{py,m}\qquad \bar{C}_{m,y}=C_{y,m}.$$

\noindent
Some care must be exercised when computing these correlators, since the use of
coherent states implies that one has a non-trivial inner product. These overlaps
can equivalently be computed as correlators in the free matrix model in accord
with the state operator map. As discussed in section 2, our transformation makes
use of these correlators. The cubic contribution from \HInt\ becomes

$$\eqalign{ H_3 &=\left[-{1\over 2} C_{n,qz} \left( 8\pi^2 n^2 - 8\pi^2
{q^2 \over z^2} - 0 \right) + D_{qz,n}\right] \psi_1^{n1{\dagger}} \psi_1^{qz}
\psi_0^{(1-z)} \cr
& + \left[ - {1\over 2} C_{n,z} \left( 8\pi^2 n^2 - 0 - 0 \right) +
D_{z,n}\right] \psi_1^{n1{\dagger}} \psi_2^z \psi_3^{(1-z)}\cr
& + \left[ + {1\over 2} \bar{C}_{m,qy} \left(0+8\pi^2 m^2 - {8q^2\pi^2\over y^2}
\right) + D_{m,qy}\right] \psi_0^{(1-y){\dagger}} \psi_1^{qy{\dagger}}
\psi_1^{m1}\cr
& + \left[ {1\over 2} \bar{C}_{m,y} \left( 8\pi m^2 - 0 - 0 \right) \right]
\psi_2^{y{\dagger}} \psi_3{(1-y)^{\dagger}} \psi_1^{m1} }
$$

\noindent
We then evaluate the coefficient in the respective couplings

$$\eqalign{
-{1\over 2} C_{n,qz} \left( 8\pi^2 n^2 -{8\pi^2 q^2\over z^2} \right)
+ D_{qz,n}
& = -4\pi^2 \left( n^2 - {q^2\over z^2} \right) {z^{3/2}\sqrt{1-z}\over
\sqrt{J}\pi^{2}} \,\,\, {\sin^2 (\pi nz )\over (q - nz)^{2}}\cr
&- {1\over\sqrt{J}}
{\sqrt{1-z\over z}} \, {8 nz\over (q-nz)} \, \sin^2 (\pi n z) \cr
& = 8\pi^2 \left( {\sqrt{1-z\over z}} \, {1\over 2\sqrt{J} \pi^{2}} \sin^2 ( \pi
n z ) \right)}$$

$$\eqalign{
& {1\over 2} \bar{C}_{m,py} \left( 8\pi^2 m^2 - {8\pi^2 p^2\over y^2} \right) +
D_{m,py} = 4 \pi^2 C_{py,m} \left( m^2 - {p^2\over y^2} \right) + D_{m,py}\cr
& = 4\pi^2 y^{3/2} {\sqrt{1-y}\over\sqrt{J} \pi^2} \,\,\, {\sin^2 (\pi my)\over
(p-my)^2} \left( m^2 - {p^2\over y^2} \right) - {1\over\sqrt{J}}\,\, {\sqrt{1-y
\over y}} \left( {8p\over my-p} \right) \sin^2 (\pi my)\cr
& = 8\pi^2\left[ {1\over 2\sqrt{J}\pi^2} \sqrt{{1-y\over y}}\, \sin^2 (\pi my
)\right],}$$

$$\eqalign{ - {1\over 2} C_{n,z}
\left( 8\pi^2n^2\right) + D_{z,n}
&=\left( -4\pi^2 n^2 \right) \left( {-\sin^2 (\pi nz)\over \pi^2 \sqrt{J} n^2
}\right)- {8\over \sqrt{J}} \sin^2 (\pi nz )\cr
&= 8\pi^2 \left( {-\sin^2 (\pi my
)\over 2\sqrt{J} \pi^2 } \right)}$$

$$ {1\over 2} C_{y,m} \left( 8\pi^2 n^2 \right) = 8\pi^2 \left( {-\sin^2 (\pi my
)\over 2\sqrt{J} \pi^2 } \right)
$$

\noindent
Collecting these couplings, we obtain

$$\eqalign{H_3&=\lambda'g_2 8\pi^2\Big(\tilde{\Gamma}^{(1)}_{n,my}
\psi_1^{n1\dagger}\psi_1^{my}\psi_0^{1-y}
+\tilde{\Gamma}^{(1)}_{my,n}\psi_0^{(1-y)\dagger}\psi_1^{my\dagger}\psi_1^{n1}
\cr
&+\tilde{\Gamma}^{(1)}_{n,y}\psi_1^{n1\dagger}\psi_2^y\psi_3^{(1-y)}
+\tilde{\Gamma}^{(1)}_{y,n}\psi_2^{y\dagger}\psi_3^{(1-y)\dagger}\psi_1^{n1}\Big
),}$$

\noindent
with

$$ \tilde{\Gamma}^{(1)}_{n,my}=\tilde{\Gamma}^{(1)}_{my,n}=\sqrt{1-y\over y}
{\sin^2 (\pi ny)\over 2\sqrt{J}\pi^2},$$

$$ \tilde{\Gamma}^{(1)}_{n,y}=\tilde{\Gamma}^{(1)}_{y,n}=-{1\over\sqrt{J}}
\sin^2 (\pi ny).$$ These are indeed the cubic couplings generated from the pp
wave SFT \rBKPS,\rCFHM,\rPSVVV,\rGMP.

\newsec{Lattice Strings and The Vertex}

We have in sect.2 and in (I) given the collective Hamiltonian corresponding to
$H_0$ and we now concentrate on $H_1$. We start with the contributions to the
cubic interaction.

In what follows we will consider loops of the form

$$\Phi_J(\{l_{i}\})=\Tr\Big(T_l\prod_{l=1}^{n}
b(l_i)A^{J\dagger}\Big),$$

\noindent
where

$$ b_l=A^{l \dagger}b^{\dagger}A^{-l\dagger}.$$

\noindent
The subscript $l$ on $b$ is understood as $l$ mod $J$, so that
we can assume this index is in the range $0\le l\le J-1$. The symbol $T_l$
orders the $b_l$ so that $l$ increases from left to
right. These loops correspond to the string states

$$\Phi_J(\{l_{i}\})\leftrightarrow
\sum_{p=0}^{J_{1}-1} \prod_{i=1}^{n_{1}} b_{p+ l_{i}mod J_{1}}^{\left(1\right)
\, \dagger}|0;J\rangle .$$

\noindent
In the continuum limit the sparse occupation of lattice sites becomes
trivial.

For simplicity,
we focus on a single complex impurity. The cubic interaction is represented as

$$
H_3^{col}=\sum_{J_1,J_2,\{l^{(1)}\},\{l^{(2)}\}} \big( H_1
\Phi_{J_1}^{n_1}(\{l^{(1)}\}) \Phi_{J_2}^{n_2}(\{l^{(2)}\}) \big)
{\partial \over \partial \Phi_{J_{1}}^{n_{1}}(\{l^{(1)}\})}{\partial \over
\partial \Phi_{J_{2}}^{n_{1}}(\{l^{(2)}\})}
$$

\noindent
where, because we have a single impurity, joining is generated by

$$\eqalign{\hat{H}_{1}= -g_{YM}^2 \Tr\left( \big[b^{\dagger},A^{\dagger} \big]
\big[{\partial\over \partial b^{\dagger}},{\partial\over \partial A^{\dagger}}
\big] \right)}.$$

\noindent
Towards this end, consider the following derivatives

$$\eqalign{{\partial \Phi_J(\{l_{i}\})\over\partial A^{\dagger}_{ij}}
&=\big(P\{l_i\}\big)_{ji}\cr &=\sum_{a=0}^{J-1} \Big[T_l \prod_{l=1}^{n}
b(l_{i}-a\, mod\, J)A^{J-1\dagger}\Big]_{ji},}$$

\noindent
and

$$\eqalign{{\partial \Phi_J(\{l_{i}\})\over\partial b^{\dagger}_{ij}}
&=\big(Q\{l_i\}\big)_{ji}\cr &=\sum_{k=1}^{n} \Big[T_l \prod_{i=1,i \neq k}^{n}
b(l_{i}-l_{k}\, mod\, J)A^{J\dagger}\Big]_{ji}.}$$

\noindent
which produce

\eqn\loopjoin {\eqalign{&\big(H_1
\Phi_{J_1}(\{l_{i}^{(1)}\})\Phi_{J_2}(\{l_{j}^{(2)}\})\big)=g_{YM}^2 \Big(\Tr
(b^{\dagger} A^{\dagger} P_1Q_2)-\Tr (A^{\dagger}b^{\dagger} P_1Q_2)
+\Tr (b^{\dagger} A^{\dagger} P_2Q_1)\cr
&-\Tr (A^{\dagger}b^{\dagger} P_2Q_1)-\Tr
(P_2b^{\dagger} A^{\dagger} Q_1)+\Tr (P_2 A^{\dagger}b^{\dagger} Q_1) -\Tr (P_1
b^{\dagger} A^{\dagger} Q_2)+\Tr (P_1 A^{\dagger}
b^{\dagger} Q_2)\Big),}}

$$\big(P_a\big)_{ji}={\partial \Phi_{J_a}(\{l^{a}_{i}\})\over\partial
A^{\dagger}_{ij}},\qquad
\big(Q_a\big)_{ji}={\partial \Phi_{J_a}(\{l^{a}_{i}\})\over\partial
b^{\dagger}_{ij}},\qquad a=1,2$$

\noindent for the loop joining terms.

In I the free matrix model was seen to generate loop joinings that were
reproducing  the string field theory interaction
vertex with the trivial energy prefactor (zeroth order in $\mu$). These free
matrix model  loop joining operations were associated with

$$
Tr \left( P_{1} \, \bar{P}_{2}\, \right) = \sum_{J,\{l\}} \, C_{J J_1
J_2}^{\{l\}\{l^{(1)}\}\{\bar{l}^{(2)}\}}\, \Phi_J(\{l\}),
$$

\noindent
and

$$
Tr \left( Q_{1} \bar{Q}_{2} \right) = \sum_{J, \{l\}} \, G_{J J_{1} J_{2}}^{\{
l\}\{l^{(1)}\}\{\bar{l}^{(2)}\}} \, \Phi_J(\{l\}).
$$

\noindent
Now from the action of \loopjoin\ we will find another sequence of joining
contributions to the full cubic interaction.
As before we may again use our correspondence but this time with a different
state from
$|V_{3}^{0}\rangle$. However, in the continuum limit (large $J$) the new state
will reduce to be $|V_{3}^{0}\rangle$ multiplied
by an insertion which will be shown to be, in combination to the effect of the
field redefinition, precisely the large $\mu$
limit of the string field theory prefactor. Consider the term

\eqn\FrstTerm {\eqalign{&\Tr (P_1 b^{\dagger} A^{\dagger} Q_2)= \cr
&\sum_{a=0}^{J_1-1} \sum_{k=1}^{n_2-1} \, Tr \left( \left(
T_l \prod_{i=1}^{n_{1}} b(l_{i}^{(1)}-a) A^{J_{1}-1 \dagger}\right) b^{\dagger}
A^{\dagger} \left( T_j \prod_{j=1,j \neq
k}^{n_{2}} \, b(l_{j}^{(2)}-l_{k}^{(2)})\right) A^{J_{2} \dagger} \right) = \cr
& \sum_{a=0}^{J_1-1} \sum_{k=1}^{n_2-1} \, Tr
\left( \left( T_l \prod_{i=1}^{n_{1}} b(l_{i}^{(1)}-a) \right) b(J_{1}-1) \left(
T_j \prod_{j=1,j \neq k}^{n_{2}} \,
b(l_{j}^{(2)}-l_{k}^{(2)}+J_{1})\right) A^{J_{1}+J_{2} \dagger} \right).}}

\noindent
The summation in $P_1$ reflects the averaging in order to have the physical
state
$|\psi_{1} \rangle$, a part of the non-triviality of the above comes from $Q_2$.
We
would expect the same to happen with the second string variables after seen in
the third string in the case of $|V_{3}^{0}\rangle$. In this case we observe the
necessity of the insertion $b^{(2)\dagger}_{J_1}$. Another factor comes from
$b^{\dagger}$ which gives an insertion $b^{(3)\dagger}_{J_1-1}$. From \FrstTerm\
we
can easily read off the appropriate state for this term to be,

$$ b^{(2) \dagger}_{J_1} b^{(3) \dagger}_{J_1-1}|V^3_0\rangle .$$

\noindent
In exactly the same way we can argue for the following identifications

$$\eqalign{\Tr (P_1 A^{\dagger} b^{\dagger} Q_2) \longrightarrow  b^{(2)
\dagger}_{J_1} b^{(3) \dagger}_{J_1}|V^3_0\rangle  .}$$

\noindent
The remaining terms follow

$$\eqalign{
\Tr (P_2b^{\dagger} A^{\dagger} Q_1)=\Tr (Q_1 P_2 b^{\dagger} A^{\dagger})
\longrightarrow & \,\,b^{(1) \dagger}_{0} b^{(3)
\dagger}_{J_1+J_2-1}|V^3_0\rangle \cr \Tr (P_2 A^{\dagger}b^{\dagger} Q_1)=\Tr
(b^{\dagger}Q_1 P_2 A^{\dagger}) \longrightarrow &
\,\, b^{(1) \dagger}_{0} b^{(3) \dagger}_{0}|V^3_0\rangle .}
$$

\noindent
Next we have the four remaining contributions. Consider

$$\eqalign{\Tr (b^{\dagger} A^{\dagger} P_1Q_2)&=\sum_{a=0}^{J_1-1}
\sum_{k=1}^{n_2-1} \,
Tr \Big( b^{\dagger} \Big( T_l
\prod_{i=1}^{n_{1}} b(l_{i}^{(1)}-a+1) \Big)\Big)\cr
&\Big(\Big(T_j \prod_{j=1,j \neq k}^{n_{2}} \,
b(l_{j}^{(2)}-l_{k}^{(2)}+J_1)\Big) A^{J_{1}+J_{2} \dagger} \Big).}$$

\noindent
After employing once more our correspondence we have

$$
b_0^{(3)\dagger} \, b_{J_{1}}^{(2)\dagger} \, |V_{3,1}^0\rangle
$$

\noindent
where

$$
|V_{3,1}^0\rangle = e^{\sum_{i=0}^{J_{1}-1} \,b_{i+1}^{(3) \dagger} \,
b_{i}^{(1)\dagger}+\sum_{j=J_{1}}^{J_{1} + J_{2} -1}
 \, b_j^{(3) \dagger} \,\, b_{j}^{(2) \dagger}} |0\rangle_{123}
$$

\noindent
Similarly

$$
\Tr (A^{\dagger}b^{\dagger} P_1 Q_2) \longrightarrow  b_1^{(3)\dagger} \,
b_{J_{1}}^{(2)\dagger} \, |V_{3,1}^0\rangle
$$

\noindent
Analogously

$$\eqalign{
\Tr (b^{\dagger} A^{\dagger} P_2Q_1)=\Tr (Q_1 b^{\dagger} A^{\dagger}
P_2)\longrightarrow & \,\,b_{J_{1}}^{(3) \dagger} \,
b_0^{(1) \dagger} \, V_{3,2}^0\cr \Tr (A^{\dagger}b^{\dagger} P_2Q_1)=\Tr (Q_1
A^{\dagger}b^{\dagger} P_2) \longrightarrow & \,\,
b_{J_{1}+1}^{(3)\dagger} \, b_0^{(1) \dagger} \, V_{3,2}^0 }
$$

\noindent
where

$$
|V_{3,2}^0\rangle = e^{\sum_{i=0}^{J_{1}-1} \,b_{i}^{(3) \dagger} \,
b_{i}^{(1)\dagger}+\sum_{j=J_{1}}^{J_{1} + J_{2} -1}
 \, b_{j+1}^{(3) \dagger} \,\, b_{j}^{(2) \dagger}} |0\rangle_{123}
$$

We are now ready to collect the contributions to the three string vertex and
take the continuum limit. Let us recall the
contribution from the free part of the Hamiltonian

$$
\left(E_3 - E_2 - E_1 \right) \vert V_3^0 \rangle
$$

\noindent This is essentially the result of I, but now we have the improved $\,
O(\lambda )\,$ corrected energies $\, E_i \,$ in
the prefactor
multiplying $\, V_3^0 \,$.  This form is deduced from our free coherent
Hamiltonian after performing the coherent physical space
map. The
relevant transformation given generally in sect.3 by eqs. \Summary-\HInt. From
the action of $\, H_1\,$ we have generated eight
terms.  In the
continuum limit, the difference between $\, V_{3,1}^0 \,$ and $\, V_3^0 \,$ is
negligible so the eight terms appear with the same
3-vertex:

$$\eqalign{
& \left[ \left( b_0^{(3)\dagger} \, - b_{J_{1} + J_{2} -1}^{(3) \dagger}
\right)\,\,   b_0^{(1) \dagger} \right. + \left(
b_{J_{1}}^{(3)
\dagger} - b_{J_{1} -1}^{(3) \dagger} \right) \, \, b_{J_{1}}^{(2) \dagger}\cr &
+ \left( b_0^{(3) \dagger} \, - b_1^{(3) \dagger}
\right) \, \,
b_{J_{1}}^{(2)\dagger } + \left.\left( b_{J_{1}}^{(3) \dagger} \, - b_{J_{1}+1}^{(3)
\dagger}\right) \,\, b_0^{(1)\dagger} \right] |V_3^0
\rangle}
$$

\noindent Altogether we therefore find the total matrix theory vertex comes with
the prefactor

$$
\hat{V}_{Matrix} = \left[ \left( E_3 - E_1 - E_2 \right) + 2 \left( \Delta
b_{0}^{(3) \dagger}-\Delta b_{J_1}^{(3) \dagger}
\right)\,\, \left(
 b_{J_1}^{(2) \dagger} - b_{0}^{(1) \dagger} \right) \right]|V_3^0 \rangle
$$

\noindent
where $\, \Delta b\,$ stands for a finite difference. We now clearly see the
leading effect of  the Y-M interaction: first it induces a correction to the
energies(dimensions) in the prefactor renormalizing the free matrix theory
result ,in
addition there are novel contributions to the prefactor .

In the continuum limit these become operator insertions at the interaction
point, in particular we get the derivative $\, b^+ (0)^1\,$ of the string creation
coordinate at the interaction point. In order to demonstrate agreement with SFT
now show that
the SFT prefactor can be written in an identical form and
$\,\hat{P}_{Matrix} \rightarrow \hat{P}_{SFT}$.

To perform the comparison consider the SFT prefactor
\rSFT

$$
\hat{P} = \sum_{r=1}^3 \sum_{n>0} {\omega_n^{(r)}\over \mu \alpha^{(r)}} \left(
a_n^{(r)\dagger}\, a_n^{(r)} -
a_{-n}^{(r)\dagger}\, a_{-n}^{(r)} \right)
$$

\noindent
We can ``separate'' out an energy contribution to write it as

$$
\hat{P} = 2 \sum_{n>0} \left\{ {\omega_n^{(1)}\over\mu\alpha^{(1)}}\,
a_n^{(1)\dagger}
 a_n^{(1)} + {\omega_n^{(2)}\over\mu\alpha^{(2)}}\,  a_n^{(2)\dagger} a_n^{(2)}
-
{\omega_n^{(3)}\over\mu\alpha^{(3)}} \, a_n^{(3)\dagger} \, a_n^{(3)} \right\} -
\left( E_1 + E_2 - E_3 \right)
$$

\noindent
where

$$E_{r} = \sum_{n} \, {\omega_n^{(r)}\over\mu\alpha^{(r)}} \, a_n^{(r)\dagger}
a_n^{(r)}
$$

\noindent
Denoting the first term containing positive modes only by $\, P_+ \,$ we have
the 3-vertex

$$
P_+ \vert V_3^0 \rangle =  2 \, \sum_{n>0} \Big\{ \left( {\omega_n^{(1)}\over
\mu \alpha^{(1)}}\, - {\omega_m^{(3)}\over\mu
\alpha^{(3)}} \right) \, a_m^{(3)\dagger} \, a_n^{(1)\dagger} \, N_{nm}^{(13)}
$$ $$+ \left( {\omega_n^{(2)}\over \mu
\alpha^{(2)}} - {\omega_m^{(3)}\over\mu\alpha^{(3)}} \right) \, a_m^{(3)\dagger}
\, a_n^{(2)\dagger} \,\, N_{nm}^{(23)} \Big\}
\vert V_3^0 \rangle
$$

\noindent
using

$$\eqalign{
& N_{nm}^{(13)} \rightarrow {2\over\pi} (-)^{m+n+1} \beta^{3/2} \,\, {m\sin
(\beta m\pi )\over m^2 \beta^2 - n^2 }\cr &
N_{nm}^{(23)}
\rightarrow  {2\over \pi} (-)^{m}( 1-\beta )^{3/2} \,\, {m\sin (\beta m\pi
)\over (1-\beta )^2 m^2 - n^2 }}
$$

\noindent
one gets

$$
P_+ \vert V_3^0 \rangle   = 2 \, {\beta^{3/2}\alpha^{(3)}\over\mu^{2}\alpha^{(1)2}\pi} \, \left( \sum_{m>0} (-)^m m \, \sin (\beta m\pi )
a_m^{(3)\dagger} \right) \left( \sum_{n>0} (-)^n a_n^{(1)\dagger} \right) - $$
$$ 2
\, {(1-\beta
)^{3/2}\alpha^{(3)}\over\mu^{2}\alpha^{(2)2} \pi} \, \left( \sum_{m>0} (-)^m m
\, \sin
(\beta m\pi ) \, a_m^{(3)\dagger} \right) \,
 \left( \sum_{n>0} \, a_n^{(2)\dagger} \right) \vert V_3^0 \rangle
$$

\noindent
since we may use the expansion

$$
b^{(3)\dagger} (\sigma ) = {1\over \mu \sqrt{2\pi\alpha_{(3)}}} \sum_{m>0} (-)^m
\left(
a_m^{(3) \dagger} \cos \left( {m\sigma \over \alpha_{(3)}} \right)
+ a_{-m}^{(3) \dagger} \sin \left( {m\sigma \over \alpha_{(3)}} \right)\right)
$$
\noindent and
$$
b^{(r)\dagger} (\sigma ) = {1\over \mu \sqrt{2\pi\alpha_{(r)}}} \sum_{m>0}
\left(
a_m^{(r) \dagger} \cos \left( {m\sigma \over \alpha_{(r)}} \right)
+ a_{-m}^{(r) \dagger} \sin \left( {m\sigma \over \alpha_{(r)}} \right)\right)
,\rm{where} \quad r=\{1,2\}
$$

\noindent
For all three strings we have:

$$
\eqalign{ -b^{(1)\dagger} (\pi
\alpha_{(1)}) & ={1\over \mu \sqrt{2\pi\alpha^{(1)}}} \sum_n \,
(-)^{n+1} a_n^{(1)\dagger}\cr b^{(2)\dagger} \left( 0 \right) & = {1\over \mu
\sqrt{2\pi\alpha^{(2)}}} \sum_n \,
a_n^{(2)\dagger}\cr {1\over2} \left( b^{(3)\dagger} ( -
\pi \alpha_1 ) - b^{(3)\dagger}\, (\pi \alpha_1) \right) & = {1\over \mu
\sqrt{2\pi\alpha^{(3)}}} \sum_m \, {m\over\alpha_3} \,
\sin (\beta m\pi ) \, a_m^{(3)\dagger}}
$$

\noindent
and

$$
P_{+} = 2 \, \left(b^{(3)\dagger\prime}
(\pi\alpha_1 ) - b^{(3)\dagger\prime} (-\pi \alpha_1)\right) \,
\left(b^{(2)\dagger}\, (0) -  b^{(1)\dagger} (\pi \alpha_1)
\right)
$$

\noindent
a form predicted in our matrix model calculation.

\newsec{Conclusions}

We have in the present paper constructed the pp wave cubic SFT interaction from
large N matrix theory using the
Berenstein-Maldacena-Nastase
limit. Even though this construction was performed at leading order in the
Yang-Mills theory coupling constant,
the  method
that we employ is generally not limited to weak coupling. We have concentrated on
the sector of the theory
generated by the Higgs
(scalar
field) degrees of freedom. This and the use of a matrix model language was  for
the purpose of notational
simplicity. The
construction can be
extended to include fermionic fields of SUSY Yang-Mills theory and also states
generated by the (covariant)
derivatives of the
Higgs fields. In
the basic scheme that we presented the closed string theory cubic interactions
are seen to be correctly
generated. Most importantly
we have seen
how the correct prefactor or operator insertion at the interaction point is
obtained from the
large $N$ matrix theory construction.
There has
been some debate on the form of the prefactor, our direct calculations provide a
unique and well
defined form. Apart from the
extension of the
present approach to include the fermionic and derivative degrees of freedom of
the full Yang-Mills
theory  the most important
future goal is
that of presenting a derivation of the full nonperturbative SFT interaction.
This implies an
extension of the calculations done
presently
without the use of weak coupling methods. The collective field approach that we
employ is in
general not limited to weak Yang-Mills
coupling,
its application to various phases of large N has been demonstrated in past
studies. For that
reason its further study offers a
possibility for
the nonperturbative understanding of the gauge theory/string theory
correspondence.

{\it Acknowledgements:}
The work of RdMK and JPR is supported by NRF grant number Gun 2053791.
The work of AD and AJ is supported by DOE grant DE FGO2/19ER40688(Task A).

$$ $$

\noindent
{\bf APPENDIX: Lattice Strings and The Spectrum}

In this section
we consider in detail the quadratic piece of the collective field theory
hamiltonian. This will be done fully with a goal of deriving the
first quantized lattice string hamiltonian of BMN. We consider the
case of  a real impurity matrix coordinates and work in the coherent state
basis. Consider the nontrivial contribution from the interacting piece of
the matrix theory. It is given by

$$
H_2^{col}=\sum_{J,\{l\}} \big( H_1\Phi_{J}(\{l\})\big)
{\partial \over \partial \Phi_{J}(\{l\})}
$$

\noindent
where in the action of $H_1 \Phi_{J}(\{l\})$ we concentrate only on
the contribution linear in the collective field $\Phi$. In the
creation annihilation operator basis, we consider the action of

$$\eqalign{-g_{YM}^2 &Tr\Big(\big[b^{\dagger},A^{\dagger}\big]
\big[{\partial\over\partial b^{\dagger}},{\partial\over\partial A^{\dagger}}
\big]+ \big[{\partial\over \partial
b^{\dagger}},A^{\dagger}\big]
\big[b^{\dagger},{\partial\over \partial A^{\dagger}}\big]+ \cr
&\big[b^{\dagger},A^{\dagger} \big] \big[b^{\dagger},
{\partial\over\partial
A^{\dagger}} \big] + \big[{\partial\over \partial b^{\dagger}},A^{\dagger} \big]
\big[{\partial\over \partial b^{\dagger}},
{\partial\over\partial A^{\dagger}} \big]\Big).}$$

\noindent
We begin by considering the most general trace

$$\Phi_{J}^{n}(\{l_i\})={1\over \sqrt{N^{J+n}}}
Tr\left(T_{l}\prod_{i=1}^n b^{n_i}(l_i) A^{\dagger J}\right)$$

\noindent
where we have $l_i \neq l_j \, \hbox{for} \, i \neq j$. The first term in our
matrix hamiltonian produces at first order in $N$,

$$
\eqalign{& -Tr\left( \big[b^{\dagger},A^{\dagger} \big] \big[{\partial\over
\partial b^{\dagger}},{\partial\over \partial
A^{\dagger}}
\big]\right) \Phi_{J}^{n}(\{l_i\})= {2N\over \sqrt{N^{J+n}}} \sum_{i=1}^n
Tr\left(T_l \prod_{i=1}^n b^{n_i}(l_i) A^{\dagger J} \right) \cr
&-{N\over
\sqrt{N^{J+n}}} \sum_{j=1}^n Tr\left(T_l \prod_{i=1}^n
b^{\delta(i,j)}(l_i-\delta(i,j)\, mod\, J)b^{n_j-\delta(i,j)}(l_i)
A^{\dagger J} \right)
\cr &-{N\over
\sqrt{N^{J+n}}} \sum_{j=1}^n Tr\left(T_l \prod_{i=1}^n
b^{\delta(i,j)}(l_i+\delta(i,j)\, mod\, J)b^{n_j-\delta(i,j)}(l_i)
A^{\dagger J} \right).}
$$

\noindent
This implies a contribution of the form

$$
H_2^{col}=\sum_{J,\{l\}} \big( \hat{h}_{col} \Phi_{J}(\{l\})\big) {\partial
\over \partial \Phi_{J}(\{l\})}
$$

\noindent
with  $\hat{h}_{col}$ being a first quantized lattice string operator. We see
from the above that we have a contribution

$$
 N \, \sum_{i=0}^{J-1} :\left(
b_{i+1}^{\dagger}b_{i+1}+b_{i}^{\dagger}b_{i}-b_{i+1}^{\dagger}b_{i}
-b_{i}^{\dagger}b_{i+1}
\right):\, .
$$

Next from the second and third term in $H_1$ we find

$$
\eqalign{&- Tr\big(\big[{\partial\over \partial b^{\dagger}},A^{\dagger} \big]
\big[b^{\dagger},{\partial\over \partial
A^{\dagger}} \big] \big)
\Phi_{J}^{n}(\{l_i\})= \cr &-{N\over \sqrt{N^{J+n}}} \sum_{j=1}^n Tr\left(T_l
\prod_{i=1}^n
b^{\delta(i,j)}(l_i-\delta(i,j)\, mod\,
J)b^{n_j-\delta(i,j)}(l_i) A^{\dagger J} \right) \cr
&-{N\over \sqrt{N^{J+n}}} \sum_{j=1}^n Tr\left(T_l
\prod_{i=1}^n
b^{\delta(i,j)}(l_i+\delta(i,j)\, mod\,
J)b^{n_j-\delta(i,j)}(l_i) A^{\dagger J} \right),}
$$

and

$$
\eqalign{-Tr\big(\big[b^{\dagger},A^{\dagger}\big]
\big[b^{\dagger},{\partial\over\partial A^{\dagger}}\big]&\big)
\Phi_{J}^{n}(\{l_i\})={N\over\sqrt{N^{J+n}}}\sum_{j=1}^n
Tr\left(T_l b^{2}(j+1\, mod\, J)\prod_{i=1}^n b^{n_i}(l_i) A^{\dagger J}
\right)\cr
&+{N\over\sqrt{N^{J+n}}} \sum_{j=1}^n Tr\left(T_l b^{2}(j)\prod_{i=1}^n
b^{n_i}(l_i) A^{\dagger J} \right) \cr
&-2{N\over \sqrt{N^{J+n}}} \sum_{j=1}^n
Tr\left(T_l b(j+1\, mod\, J) \, b(j) \prod_{i=1}^n b^{n_i}(l_i)
A^{\dagger J} \right)}
$$

\noindent
with the last term giving

$$
\eqalign{-Tr\big(\big[{\partial\over \partial b^{\dagger}},A^{\dagger} \big]
&\big[{\partial\over \partial
b^{\dagger}},{\partial\over
\partial A^{\dagger}} \big]\big) \Phi_{J}^{n}(\{l_i\})= \cr &{N\over
\sqrt{N^{J+n}}} \sum_{j=1}^n
Tr\left(T_l \theta(n_j-1)\prod_{i=1}^n b^{n_i-2\delta(i,j)}(l_i) A^{\dagger J}
\right) \cr
& +{N\over \sqrt{N^{J+n}}} \sum_{j=1}^n
Tr\left(T_l
\theta(n_{j+1}-1)\prod_{i=1}^n b^{n_{i+1}-2\delta(i,j)}(l_{i+1}) A^{\dagger J}
\right) \cr
& -2{N\over \sqrt{N^{J+n}}} \sum_{j=1}^n
Tr\left(T_l
\theta(n_j) \, \theta(n_{j+1}) \prod_{i=1}^n
b^{n_i-\delta(i,j)-\delta(i,j+1)}(l_i) A^{\dagger J} \right)}
$$

\noindent
Collecting all contribution we obtain the first quantized lattice hamiltonian of
the form

$$
\eqalign{h_{col}&=g_{YM}^2 N \, \sum_{i=0}^{J-1}
\Big(b_{i+1}^{\dagger2}+b_{i}^{\dagger2}-2b_i^\dagger b_{i+1}^\dagger
+b_{i+1}^{2}+b_{i}^{2}-2b_i b_{i+1}\cr
&\quad +b_{i+1}^{\dagger}b_{i+1}+b_{i}^{\dagger}b_{i}
-2b_{i+1}^{\dagger}b_{i}-2b_{i}^{\dagger}b_{i+1}
\Big).}
$$

\noindent
which is precisely what someone would get from h by neglecting the constant
coming from $b_{i+1}b_{i+1}^{\dagger}$ and
$b_{i}b_{i}^{\dagger}$. This is recognized as the lattice BMN string hamiltonian

$$
h_{BMN}=g_{YM}^2 N \, \sum_{i=0}^{J-1} \left(
b_{i+1}^{\dagger}+b_{i+1}-b_{i}^{\dagger}-b_{i}\right)^{2}= {1\over \epsilon^2}
\,
\sum_{i=0}^{J-1} \left(
b_{i+1}^{\dagger}+b_{i+1}-b_{i}^{\dagger}-b_{i}\right)^{2}
$$

\noindent
with the understanding that as
pointed out in\rBMN the creation-annihilation operators  are to be of Cuntz
type. The physical basis behind these oscillators
is the fact that the lattice sites should be sparsely occupied (i.e. not more
than one oscillator at a lattice site).

\listrefs
\vfill\eject
\bye